\documentclass[Physsubmission, Phys]{SciPost}

\hypersetup{
    colorlinks,
    linkcolor={red!50!black},
    citecolor={blue!50!black},
    urlcolor={blue!80!black}
}

\usepackage[bitstream-charter]{mathdesign}
\DeclareSymbolFont{usualmathcal}{OMS}{cmsy}{m}{n}
\DeclareSymbolFontAlphabet{\mathcal}{usualmathcal}

\begin{document}

\begin{center}{\Large \textbf{
On the Discrepancy between the FOPT and CIPT Approaches for Hadronic Tau Spectral Function Moments\\
}}\end{center}

\begin{center}
Andr\'e H. Hoang\textsuperscript{1,2$\star$} and
Christoph Regner\textsuperscript{1} 
\end{center}

\begin{center}
{\bf 1} University of Vienna, Faculty of Physics, Boltzmanngasse 5, A-1090 Wien, Austria
\\
{\bf 2} Erwin Schr\"odinger International Institute for Mathematics and Physics,
University of Vienna, Boltzmanngasse 9, A-1090 Wien, Austria
\\
* andre.hoang@univie.ac.at
\end{center}

\begin{center}
\today
\end{center}


\definecolor{palegray}{gray}{0.95}
\begin{center}
\colorbox{palegray}{
  \begin{tabular}{rr}
  \begin{minipage}{0.1\textwidth}
    \includegraphics[width=35mm]{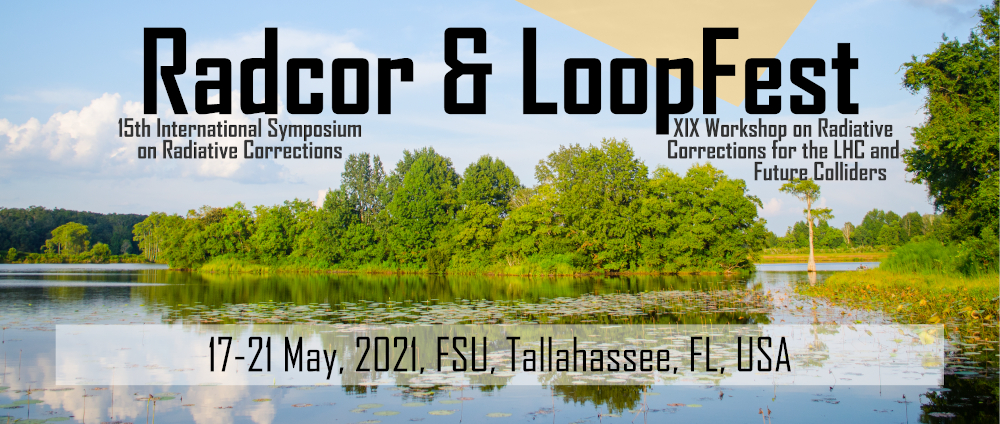}
  \end{minipage}
  &
  \begin{minipage}{0.85\textwidth}
    \begin{center}
    {\it 15th International Symposium on Radiative Corrections: \\Applications of Quantum Field Theory to Phenomenology,}\\
    {\it FSU, Tallahasse, FL, USA, 17-21 May 2021} \\
    \doi{10.21468/SciPostPhysProc.?}\\
    \end{center}
  \end{minipage}
\end{tabular}
}
\end{center}

\section*{Abstract}
{\bf
The discrepancy between the FOPT and CIPT approaches for hadronic $\tau$ spectral function moments constitutes the major theoretical uncertainty
for strong coupling determinations from tau decay data. We show the discrepancy can be analytically 
understood since the Borel representations -- which have been assumed to be identical for both approaches previously --
differ in the presence of IR renormalons. This implies that the OPE condensate corrections are different for both
approaches and that the discrepancy may eventually be reconciled. 
In the talk we explain the difference and some mathematical aspects of of the FOPT and CIPT Borel representations and show numerical results.
}

\vspace{10pt}
\noindent\rule{\textwidth}{1pt}
\tableofcontents\thispagestyle{fancy}
\noindent\rule{\textwidth}{1pt}
\vspace{10pt}

\section{Introduction}
\label{sec:intro}

Moments of the $\tau$ hadronic spectral functions obtained from LEP~\cite{Davier:2013sfa,Ackerstaff:1998yj} provide an important tool for precise determinations of the strong coupling $\alpha_s$. Predictions for the spectral function moments in the massless quark limit are based on the vacuum polarization function
$\Pi(p^2)$, which is known perturbatively to 5 loops (i.e.\ ${\cal O}(\alpha_s^4)$) in full QCD~\cite{Gorishnii:1990vf,Surguladze:1990tg,Baikov:2008jh,Herzog:2017dtz}.
Considering only for first generation quarks and QCD corrections, the theoretical moments can be written as~\cite{Braaten:1991qm,LeDiberder:1992jjr,Boito:2014sta,Pich:2016bdg}
\begin{equation}
\label{eq:momdef}
A_{W}(s_0) \, =\, \textstyle \frac{N_c}{2} \,S_{\rm ew}\,|V_{ud}|^2 \Big[\,
\delta^{\rm tree}_{W} + \delta^{(0)}_{W}(s_0)  +
\sum_{d\geq 4}\delta^{(d)}_{W}(s_0) \Big] \,,
\end{equation}
where $N_c=3$, $S_{\rm ew}$ stands for electroweak corrections (which we do not consider further), $V_{ud}$ is a CKM matrix element and $s_0$ is the upper bound of the spectral function integration.
The term $\delta^{\rm tree}_{W}$ is the tree-level contribution and $\delta^{(0)}_{W}(s_0)$ stands for the higher order perturbative QCD corrections. The terms $\delta^{(d)}_{W}(s_0)$ represent condensate corrections in the framework of the operator product expansion (OPE)~\cite{Shifman:1978bx}. They involve vacuum matrix elements of low-energy QCD operators of increasing dimension resulting from an expansion in inverse powers of $s_0$. The leading dimension $d=4$ term is related to the well-known gluon condensate $\langle \alpha_s G^{\mu\nu}G_{\mu\nu}\rangle$. There are also so-called duality-violation corrections which can be important phenomenologically, but which are not relevant for the subsequent discussion and therefore suppressed in Eq.(\ref{eq:momdef}). 
Using the 5-loop results~\cite{Gorishnii:1990vf,Surguladze:1990tg,Baikov:2008jh,Herzog:2017dtz} an impressive precision of about $5\%$ has been achieved for $\alpha_s(m_\tau^2)$ (corresponding to an uncertainty of $1.5\%$ for $\alpha_s(m_Z^2)$), where the uncertainty is dominated by the perturbative error in $\delta^{(0)}_{W}(s_0)$~\cite{Boito:2014sta,Pich:2016bdg,Boito:2020xli,ParticleDataGroup:2020ssz}.

The QCD corrections  $\delta^{(0)}_{W}(s_0)$ are obtained from the expression ($x\equiv s/s_0$)
\begin{equation}
\label{eq:deltadef}
\textstyle \delta^{(0)}_{W}(s_0)\, =\,\frac{1}{2\pi i}\,\, 
\ointctrclockwise_{{\cal C}_s}\!\! \frac{{\rm d}s}{s}\,W({\textstyle \frac{s}{s_0}})\,\hat D(s)
\, =\,
\frac{1}{2\pi i}\,\, \ointctrclockwise_{{\cal C}_x}\!\! \frac{{\rm d}x}{x}\,W(x)\,\hat D(x s_0)\,.
\end{equation}
where $\hat D(s)$ is the partonic Adler function, 
$\frac{1}{4\pi^2}(1+\hat D(s)) \, \equiv \, -\,s\,\frac{{\rm d}\hat\Pi(s)}{{\rm d} s}$ and the weight function $W(x)$ is a polynomial in $x$ which (together with the choice of $s_0$) specifies the type of moment considered. The contour path ${\cal C}_s$ (${\cal C}_x$) starts/ends at $s=s_0\pm i 0$ ($x=1\pm i0$) and traverses the complex $s$-plane, crossing the Euclidean axis half way through, with sufficient distance from the origin such that the strong coupling stays in the perturbative regime. Through analyticity this path is related to an associated integration along the real positive $s$-axis over the experimental spectral function data~\cite{Pich:2016bdg}. Frequently a circular path with $|s|=s_0$ ($|x|=1$) is considered, but it may be deformed arbitrarily as long as it stays in the region where the strong coupling remains perturbative.
For $W_\tau(x)=(1-x)^3(1+x)=1-2x+2x^3-x^4$ and $s_0=m_\tau^2$ the moment $A_{W_\tau}(m_\tau^2)$ agrees with the normalized total hadronic $\tau$ decay rate  $R_\tau=\Gamma(\tau^-\to\mbox{hadrons}\,\nu_\tau(\gamma))/\Gamma(\tau^-\to e^-\bar{\nu}_e\nu_\tau(\gamma))$. 

The two widely employed methods to calculate $\delta^{(0)}_{W}(s_0)$ are Fixed
Order Perturbation Theory (FOPT) and Contour Improved Perturbation Theory (CIPT). The CIPT approach is based on the perturbative series for the Adler function of the form\footnote{Here we use conventions, where 1-loop $\beta$-function coefficient has the form  $\beta_0=11-2n_f/3$ and we furthermore define
$a(-x) \equiv \frac{\beta_0\,\alpha_s(-s)}{4\pi} = \frac{\beta_0\,\alpha_s(-x s_0)}{4\pi}$ and $a_0\equiv \frac{\beta_0\,\alpha_s(s_0)}{4\pi} =  a(1)$. We also take $n_f=3$.
}
\begin{eqnarray}
\label{eq:AdlerseriesCIPT} 
\hat D(s) & \, =  \,& 
\textstyle \sum_{n=1}^\infty \, 
\bar c_{n} \,a^n(-x)\,, 
\end{eqnarray}
with real-valued coefficients $\bar c_{n}$ (which agree with those of the real-valued Euclidean Adler function for $s=-s_0$) and complex-valued powers of the strong coupling. One carries out the contour integration over powers of the complex-valued strong coupling $\alpha_s(-s)$. The CIPT series arises from truncating the sum in Eq.\,(\ref{eq:AdlerseriesCIPT}).
The FOPT approach consists of expanding the series~(\ref{eq:AdlerseriesCIPT}) in powers of $\alpha_s(s_0)$, so that the complex phases appear exclusively in powers of $\ln(-s/s_0)$ within the integrands of the coefficients. For FOPT, the series arises from truncating the sum in powers of $\alpha_s(s_0)$, so that the powers of the strong coupling can be factored out of the contour integration for each series term. The CIPT approach differs from  FOPT in that it resums the powers of $\ln(-s/s_0)$ to all orders along the integration path~\cite{LeDiberder:1992jjr,Pivovarov:1991rh}. 

It is an important fact that, after the contour integration~(\ref{eq:deltadef}) is carried out for the moment series, it is not possible anymore to switch between the FOPT and CIPT expansions by a scheme change of $\alpha_s$. So the difference of the truncated FOPT and CIPT series for the spectral function moments, is of a quite different character as the renormalization scale variations usually carried out for perturbative QCD predictions. 
A major limitation of $\alpha_s$ determinations from the moments $A_{W}(s_0)$ is that FOPT and CIPT calculations of  $\delta^{(0)}_{W}(s_0)$ for moments with good perturbative convergence yield to systematic numerical differences that do not seem to be covered by the conventional perturbative uncertainty estimates related to renormalization scale variations. Since CIPT in general leads to smaller values for $\delta^{(0)}_{W}(s_0)$ than FOPT, extractions of $\alpha_s(m_\tau^2)$ based on CIPT generally arrive at larger values than those based on FOPT. 

\section{Essence of this talk}
\label{sec:essence}

In this talk we report on the results given in Refs.~\cite{Hoang:2020mkw,Hoang:2021nlz}, which demonstrated that the different character of the FOPT and CIPT spectral function moment series together with the fact the coefficients $\bar c_{n}$ in Eq.~(\ref{eq:AdlerseriesCIPT}) contain asymptotic (i.e.\ non-convergent) contributions due to infrared (IR) renormalons~\cite{Mueller:1984vh,Beneke:1998ui} leads to a systematic disparity in the high-order behavior of the two types of moment series. 
It is the above mentioned property of the FOPT and CIPT methods -- that one cannot switch between them through a change of renormalization scheme -- that is a crucial ingredient in the discussions that follow.   
The disparity -- which we call the \textbf{asymptotic separation} -- can be sufficiently sizeable and manifest itself already at very low orders to explain the observed discrepancy between 5-loop FOPT and CIPT moments mentioned above. However, the disparity provides a resolution to the FOPT-CIPT discrepancy problem only if the asymptotic character is already manifest in the known perturbative coefficients up to 5-loops, which means that the known 5-loop coefficient of $\hat D(s)$ is already dominated in a sizeable way by the asymptotic behavior of infrared (IR) renormalons.
In practice, the IR renormalon dominance assumption implies that the dominant gluon condensate IR renormalon governs the behavior of the Adler function series at 5-loops~\cite{Beneke:2008ad} in a sizeable way and that -- within some uncertainties -- one can make relatively rigid predictions for the Adler function's perturbative coefficients beyond 5 loops using the renormalon calculus.
Some evidence has been provided supporting the IR renormalon dominance assumption for $\hat D(s)$~\cite{Beneke:2012vb}, but we stress that it cannot be strictly proven. Thus, even though the disparity between the FOPT and CIPT series exists as a matter of principle (because of the existence of IR renormalons in perturbative QCD~\cite{Mueller:1984vh,Beneke:1998ui}) our results
provide an explanation of the observed FOPT-CIPT discrepancy at the 5 loop level only in the context of the IR renormalon dominance assumption. We stress that this talk is not intended to provide arguments on the validity of the IR renormalon dominance assumption, but to discuss the principal aspects of the asymptotic separation. 

In the following we provide a brief primer to the renormalon calculus (Sec.~\ref{sec:primer}), explain how the character of the FOPT and CIPT series leads to principal differences in their Borel representation in the presence of IR renormalons (Sec.~\ref{sec:Borelrepresentations}), we address some mathematical subtleties of the CIPT Borel representation, and we show numerical results (Sec.~\ref{sec:implications}). We emphasize that the study of the involved analytic expressions is a complicated matter, particularly in full QCD, so that in this talk we can primarily state the outcome without going into technical details. Many analytic results will for simplicity be written down in the large-$\beta_0$ approximation (see Ref.~\cite{Hoang:2021nlz}). We refer to Ref.~\cite{Hoang:2020mkw} for all details and the analytic results in full QCD.

\section{Brief Primer on Renormalon Calculus}
\label{sec:primer}

The renormalon calculus provides a convenient way to quantify the large-order behavior of the coefficients of asymptotic series, which for any perturbative series in QCD is tied to the IR and UV properties of the $\beta$-function~\cite{Gross:1974jv,tHooft:1977xjm,David:1983gz,Mueller:1984vh,Beneke:1998ui}. Furthermore, there is a one-to-one correspondence of the asymptotic contributions in the series coefficients with IR origin to power corrections in the context of the OPE. In the following we briefly outline the basics of renormalon calculus to the extend needed for the understanding of the following parts of this talk. 

Starting from the perturbation series $\hat\sigma=\sum_{n=1}^{\infty}d_n(\mu) (\alpha_s(\mu)/\pi)^n$ for a quantity $\sigma$ in powers of (the real-valued) $\alpha_s(\mu)$, the so-called \textbf{Borel function} (or Borel transform) of $\hat\sigma$, is defined by $B[\hat\sigma](u)=\sum_{n=1}^{\infty}(4^n d_n(\mu))/(\beta_0^n \Gamma(n))u^{n-1}$. In the Borel function the asymptotic $n$-factorial growth of the $d_n$ coefficients with $n$ is compensated by the inverse powers of $\Gamma(n)$ such that the Taylor series for $B[\hat\sigma](u)$ in powers of $u$ is absolute convergent in a circle around the origin of the complex Borel $u$-plane. The resummed function $B[\hat\sigma](u)$ in this circle can be analytically continued into the entire Borel plane (at least as far as information accessible to perturbation theory is concerned). The original series (in powers of $\alpha_s$) can be recovered from the $B[\hat\sigma](u)$ Taylor series from the relation $\hat\sigma=\int_0^\infty du \,B[\hat\sigma](u)\,e^{-4\pi u/(\alpha_s(\mu)\beta_0)}$. 
The so-called \textbf{Borel sum} is the result of the same integral using the full function $B[\hat\sigma](u)$ in the entire complex $u$ plane. We call the integral over the full  $B[\hat\sigma](u)$ function also the \textbf{Borel representation} of $\sigma$. Asymptotic contributions in the original series are related to non-analytic structures (cuts and poles) in  $B[\hat\sigma](u)$ in the complex $u$ plane, where the previously mentioned radius of convergence is related to the non-analytic structure located closest to the origin. The closer the non-analytic structure is to the origin, the larger its impact (i.e.\ its dominance) in the original series. One calls these non-analytic structures \textbf{renormalons}, and one furthermore distinguishes between IR and UV renormalons. The character of these renormalons is determined from the UV and IR properties of QCD which are directly tied to the perturbative $\beta$-function and, as far as IR renormalons are concerned, to the form of the OPE corrections. One can consider the Borel sum as the ``all-order resummed'' result of the original series for $\sigma$. However, if there are non-analytic structures along the positive real $u$-axis, which usually happens for IR renormalons, the Borel sum requires some path deformation prescription, such as the principle value (PV) prescription, to be well-defined. Carrying out a convergent scheme change for $\alpha_s$ (e.g.\ related to a reexpansion of the series for a different renormalization scale or when using a different renormalization condition for the strong coupling) leaves the Borel sum invariant. We also mention that the Borel representation (and its value within any prescription) is strictly invariant under a rescaling of the coupling constant, $\alpha_s(\mu)\to \eta(\mu) \equiv \lambda\alpha_s(\mu)$ for any \textbf{positive real} number $\lambda$. The latter invariance  will be important in this talk.  

The Borel function $B[\hat D](u)$ for the perturbative Euclidean Adler function series with the form $\hat D(-s_0)=\sum_{n=1}^\infty \bar c_{n}a^n_0$ has been studied intensely in the past~\cite{Beneke:1998ui,Jamin:2005ip,Beneke:2008ad,Caprini:2009vf,DescotesGenon:2010cr,Caprini:2011ya,Beneke:2012vb}. The exact form of $B[\hat D](u)$ is unknown, but each OPE term implies the existence to an additive contribution in  $B[\hat D](u)$ with a specific non-analytic structure that is uniquely tied to the dimension of the non-perturbative matrix element (condensate), its anomalous dimension, its Wilson coefficient and the coefficients of the $\beta$-function. The leading OPE term is the dimension-4 gluon condensate correction, which for the Euclidean Adler function has the form
\begin{eqnarray}
\label{eq:gluoncondensate} 
\hat D^{\rm OPE}_{d=4}(-s_0) & \, =  \,&  \textstyle
C_{G^2}(\alpha_s(s_0))\frac{\langle \alpha_s G^{\mu\nu}G_{\mu\nu}\rangle}{s_0^2}\,.
\end{eqnarray}
The gluon condensate correction implies the existence in $B[\hat D](u)$ of a certain linear combination of non-analytic terms $\frac{1}{(2-u)^\gamma}$ for different rational values of $\gamma$. In the large-$\beta_0$ approximation, where $C_{G^2}=1$, this linear combination collapses to the single term $\frac{1}{2-u}$. In general, OPE condensate corrections with dimension $d$ are associated to certain linear combinations of non-analytic terms $\frac{1}{(d/2-u)^\gamma}$, each of which then implies contributions in the coefficients $d_n$ of the form $(\frac{\beta_0}{2d})^n(\frac{2}{d})^{\gamma-1}\frac{\Gamma(\gamma+n-1)}{\Gamma(\gamma)}$. This reduces to $(\frac{\beta_0}{2d})^n\Gamma(n)$ for $\gamma=1$ and makes the asymptotic character of the series expansion manifest. The smaller the dimension $d$, the stronger is the increase with $n$. 
Since OPE corrections are known to exist for all integer values of $d\ge 4$, $B[\hat D](u)$ contain non-analytic renormalon terms of the form $\frac{1}{(p-u)^\gamma}$ for $p=2,3,4\ldots$.
The practical limitation of the association of an OPE correction and a specific linear combination of non-analytic renormalon contributions is that the normalization of this linear combination within $B[\hat D](u)$ is a priori unknown (up to the fact that it is non-zero) and can only be fixed with additional assumptions. This is the origin of the issue concerning the renormalon dominance assumption mentioned in Sec.~\ref{sec:essence}. Only for the large-$\beta_0$ approximation, which can be calculated from massless fermion self-energy insertions into the ${\cal O}(
\alpha_s)$ gluon exchange diagrams, the Borel function is known exactly~\cite{Broadhurst:1992si}. The large-$\beta_0$ approximation is believed to exhibit at least the qualitative features of the Borel function in full QCD. 

We also mention that the gluon condensate OPE corrections almost completely cancel (up to contributions coming to the higher order corrections to its Wilson coefficient $C_{G^2}$) from the contour integral of Eq.~(\ref{eq:deltadef}), if the weight function $W(x)$ does not contain a quadratic term $x^2$. At the same time, the associated perturbative behavior of $\delta^{(0)}_{W}(s_0)$ is much better than for weight functions with a quadratic term~\cite{Beneke:2012vb}. This is the reason why for most recent phenomenological analyses (aiming for strong coupling determinations) only moments with weight functions without a quadratic term have been employed. This observation is consistent with the norm of the gluon condensate renormalon being quite sizeable, so that it already governs the size of the known 5-loop corrections.

\section{The FOPT and CIPT Borel Representations}
\label{sec:Borelrepresentations}

The central aspect of our work is that the Borel representations of the FOPT and CIPT spectral function moment series are not identical. To see this we construct the two Borel representations directly from the series terms using the form of the Borel function of the Euclidean Adler function $B[\hat D]$ as an input, but making no further assumption about their form.

We first consider CIPT and start from the observation that the contour integrals over $\frac{1}{x} W(x)(\frac{\alpha_s(-x s_0)}{\pi})^n$, which arise for each CIPT moment series term, do a priori not allow to cleanly identify the expansion parameter of the series -- simply because the renormalization scale of $\alpha_s$ is integration parameter dependent. This is the very special characteristics of the CIPT approach. It implies that we should consider the whole integral to be part of the series coefficients and reintroduce an expansion parameter by hand so that we can apply the principles in the construction of the Borel function explained in Sec.~\ref{sec:primer}.
Applying the strict invariance mentioned at the very end of the 2nd paragraph, an appropriate choice of the expansion parameter is
$\alpha_s(s_0)$, which one can conveniently pull out of the series coefficients with the appropriate power,\footnote{Any multiple of $\alpha_s(s_0)$ can be picked as the expansion parameter without changing the series terms, but $\alpha_s(s_0)$ is convenient since it also used for the FOPT series and allows for easy comparison.}
\begin{equation}
\label{eq:deltaCIPT2}
\textstyle 
\delta^{(0),{\rm CIPT}}_{W}(s_0)\, =\, 
\,\frac{1}{2\pi i} \,  \sum_{n=1}^\infty \bar c_{n} \,\Big[\,
\ointctrclockwise_{{\cal C}_x}\!\! \frac{{\rm d}x}{x}\,W(x)\,\Big({\textstyle \frac{a(-x)}{a_0}}\Big)^n\,\Big]\,
a_0^n\,.
\end{equation}
Now we can proceed and obtain the Borel function for the CIPT series $\delta_{W}^{(0),{\rm CIPT}}(s_0)$,
\begin{eqnarray}
\label{eq:BorelCIPTTaylor}
B[\delta_{W}^{(0),{\rm CIPT}}(s_0)](u) & = & \textstyle
\sum_{n=1}^\infty  \,\Big[\,  \frac{1}{2\pi i}\,\ointctrclockwise_{{\cal C}_x} \frac{{\rm d}x}{x} \, W(x) \,
\Big({\textstyle \frac{a(-x)}{a_0}}\Big)^n
\,\Big]\,\frac{\bar c_{n}}{\Gamma(n)}\,\bar u^{n-1} \\ \nonumber
& = & \textstyle
\frac{1}{2\pi i}\,\ointctrclockwise_{{\cal C}_x} \frac{{\rm d}x}{x} \, W(x)\, \Big({\textstyle \frac{a(-x)}{a_0}}\Big)\,
B[\hat D]\Big(\textstyle\frac{a(-x)}{a_0}\bar u\Big)\,,
\end{eqnarray}
where $B[\hat D](u)$ is the Borel function of the Euclidean perturbative Adler function already mentioned above, defined through the series $B[\hat D](u)=\sum_{n=1}^\infty \frac{\bar c_{n}}{\Gamma(n)}u^{n-1}$ in its region of convergence around the origin. 
The non-analytic structures in the analytically continued expression for $B[\hat D](u)$ in the entire complex $u$-plane are inherited directly to $B[\delta_{W}^{(0),{\rm CIPT}}(s_0)](u)$. The Borel representation of the spectral function moments in the CIPT approach thus has the form
\begin{equation}
\label{eq:BorelCIPT}
\textstyle
\delta_{W,{\rm Borel}}^{(0),{\rm CIPT}}(s_0) = \int_0^\infty \!\! {\rm d} \bar u \,\,  
\frac{1}{2\pi i}\,\ointctrclockwise_{{\cal C}_x} \frac{{\rm d}x}{x} \, W(x) \,
\big({\textstyle \frac{a(-x)}{a_0}}\big)\,
B[\hat D]\Big({\textstyle \frac{a(-x)}{a_0}} \bar u\Big)
\,e^{-\frac{\bar u}{a_0}}\,.
\end{equation}	
This derivation does not depend on a particular form of $B[\hat D](u)$. It only assumes that the Taylor series for $B[\hat D](u)$ in the complex $u$ plane around the origin specifies the function unambiguously in the entire complex $u$-plane. This is an assumption that has been made in any past study of the Borel function of the Euclidean perturbative Adler function, even if not explicitly stated. Furthermore, it is assumed that swapping the $x$-integration and the sum over $n$ in Eq.~(\ref{eq:BorelCIPTTaylor}), which is correct within the radius of convergence, is also allowed for the analytically continued function. We also note that it is allowed to swap the $\bar u$ and $x$ integrations when evaluating Eq.~(\ref{eq:BorelCIPT}).  

Let us now consider the Borel representation for the perturbative moments in the FOPT approach. To derive the Borel representation from the prescription in the 2nd paragraph of Sec.~\ref{sec:primer} is not an easy task due to the appearance of the powers of logarithms $\ln(-s/s_0)$ in the integrals for the series coefficients, which depend on the $\beta$-function coefficients. The derivation is, however, straightforward in the large-$\beta_0$ approximation, where the series for the Adler function in the complex plane for the expansion in powers of $\alpha_s(s_0)$ can be written down in closed form (using $a(-x)=\frac{a_0}{1+a_0\ln(-x)}$)
\begin{eqnarray}
\label{eq:AdlerseriesFOPT}
\hat D(s) & \, =  \,& \,  \textstyle
\sum_{n=1}^\infty\,
a_0^n \, \sum_{i=0}^{n-1} \frac{(n-1)!}{i!(n-i-1)!}\,\bar c_{n-i}\,(-\ln(-x))^{i} \,.
\end{eqnarray}
It is straightforward to show through algebraic manipulation that the Borel function of the resulting FOPT moment series has the form (see Ref.~\cite{Hoang:2021nlz})
\begin{equation}
\label{eq:BorelfctFOPT}
\textstyle
B[\delta_{W}^{(0),{\rm FOPT}}(s_0)](u)  =  \frac{1}{2\pi i}\,\ointctrclockwise_{{\cal C}_x} \frac{{\rm d}x}{x} \, W(x)\,  B[\hat D](u) \,  e^{-u\ln(-x)}\,,
\end{equation} 
where we again assume that swapping the contour integration and the sum over $n$, which is correct within the radius of convergence, is also allowed for the analytically continued function.\footnote{Note that the radius of convergence of the $\bar u$-series for the CIPT moment Borel function in Eq.~(\ref{eq:BorelCIPTTaylor}) is by a factor of $\alpha_s(s_0)/\alpha_s(-s_0)$ larger than the one for the $u$-series for the FOPT moment Borel function in Eq.~(\ref{eq:BorelfctFOPT}).} Using the relation $e^{-u\ln(-x)}=e^{-\frac{u}{a(-x)}+\frac{u}{a_0}}$ we then obtain the expression 
\begin{equation}
\label{eq:BorelFOPT}
\textstyle
\delta_{W,{\rm Borel}}^{(0),{\rm FOPT}}(s_0) = {\rm PV} \int_0^\infty \!\! {\rm d} u \,\, 
\frac{1}{2\pi i}\,\ointctrclockwise_{{\cal C}_x} \frac{{\rm d}x}{x} \, W(x) \,
B[\hat D](u)\,e^{-\frac{u}{a(-x)}}\,,
\end{equation}
for the Borel representation of the spectral function moments in the FOPT approach. As for Eq.~(\ref{eq:BorelCIPT}), it is allowed to swap the $u$ and $x$ integrations  when evaluating Eq.~(\ref{eq:BorelFOPT}).   It has been shown in Ref.~\cite{Hoang:2020mkw} that this expression also applies in full QCD. Prior to our work, the expression in Eq.~(\ref{eq:BorelfctFOPT}) has been adopted as the Borel representation for the FOPT moments and the CIPT moments, where the apparent relation to the CIPT expansion was taken for granted using the argumentation that its $u$ expansion immediately leads to the CIPT series of Eq.~(\ref{eq:deltaCIPT2}). However, this view did not properly account for the fact that $\alpha_s(-s)$ cannot be used as the expansion parameter of the CIPT moments series from the mathematical perspective.

\begin{figure} 
	\centering
    \includegraphics[width= 0.49\textwidth]{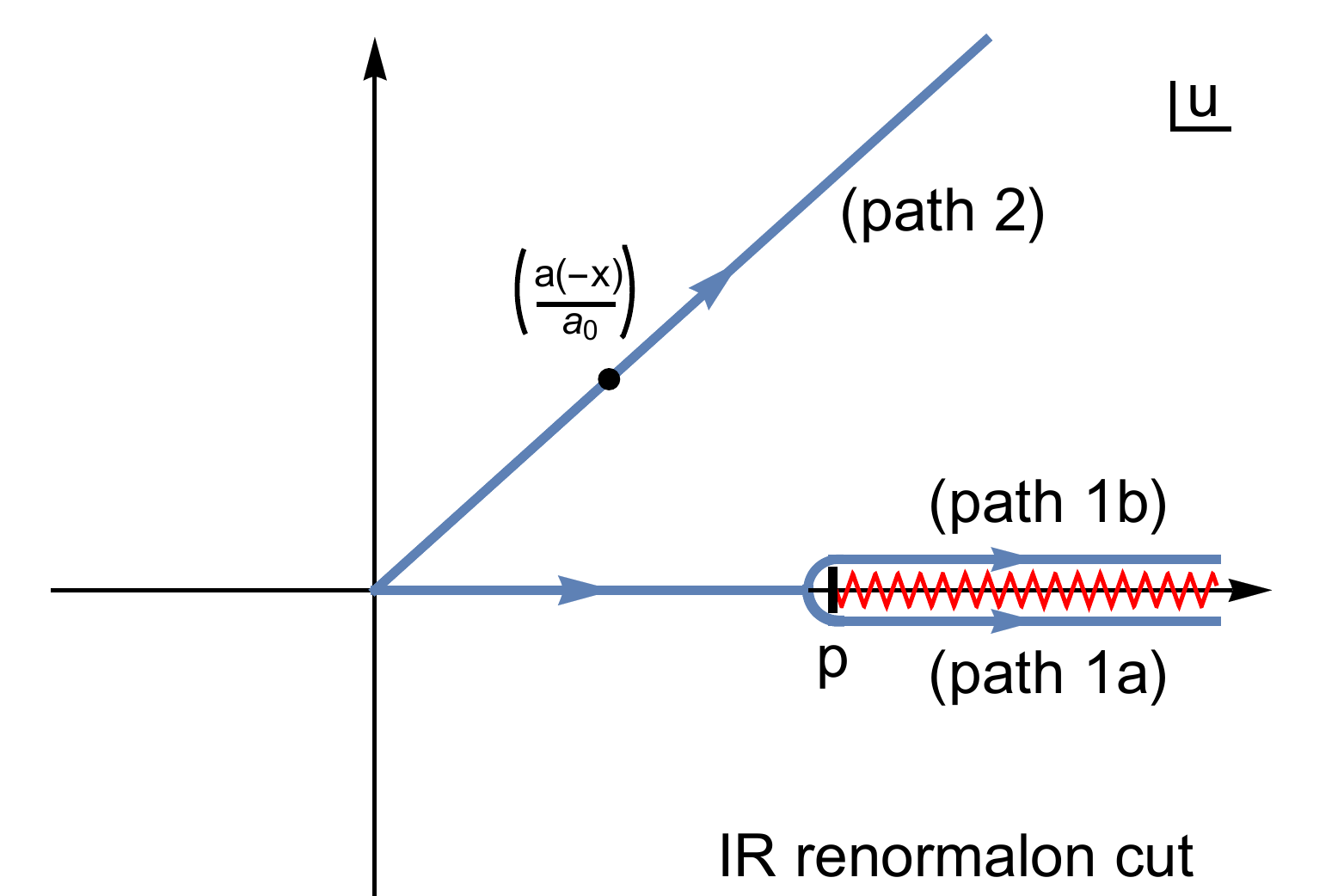}
     \includegraphics[width= 0.49\textwidth]{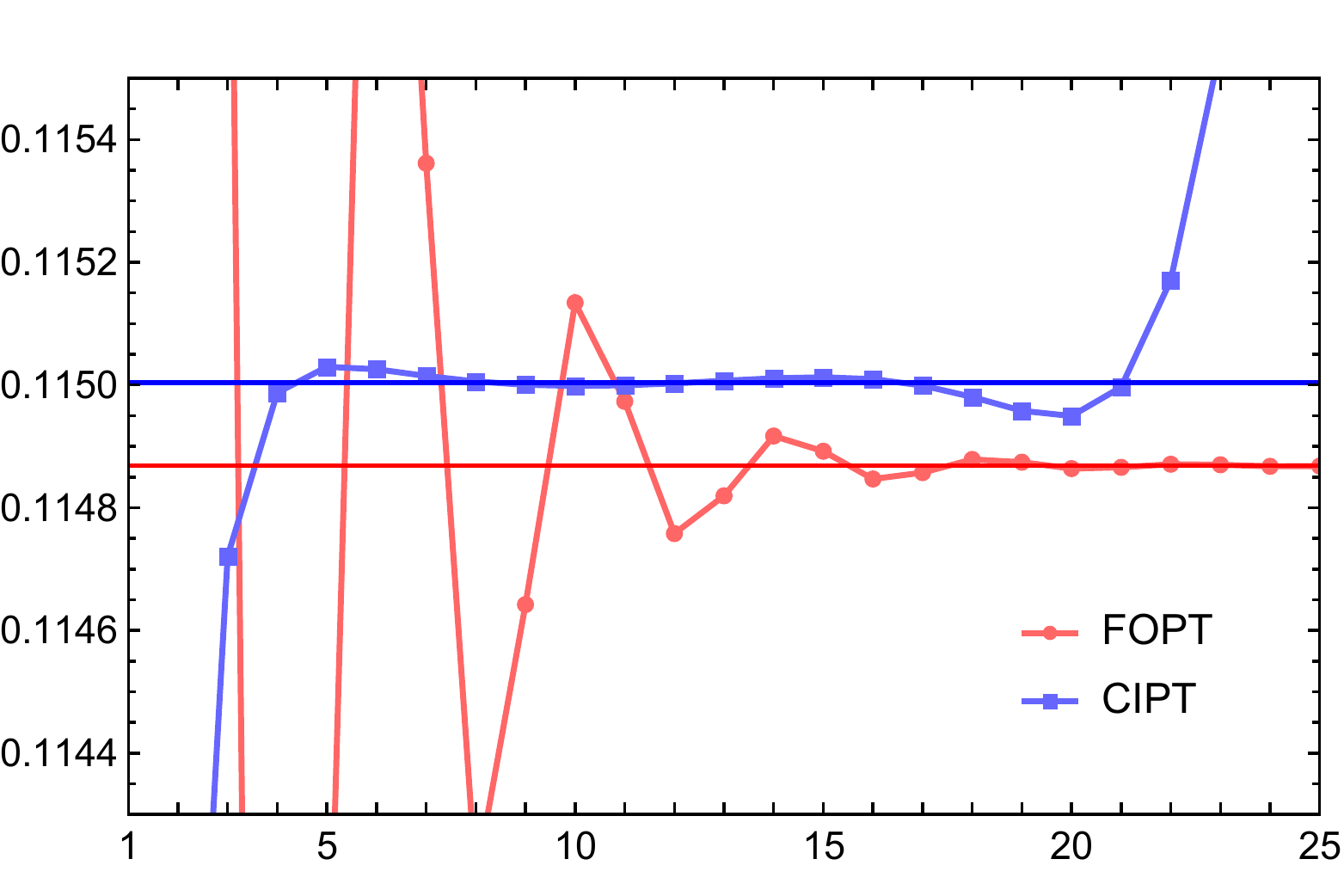}
	\caption{\label{fig:ucontour} 
		Left: Borel integration paths in the $u$-plane involved for the FOPT and CIPT Borel representations for an IR renormalon with $p>0$. The red zig-zag line represents the renormalon cut. Right: FOPT and CIPT spectral function moment series associated to the weight function $W(x)=1$ and the Borel function $\frac{1}{2-u}$ in the large-$\beta_0$ approximation for $\alpha_s(m_\tau^2)=0.34$ and $s_0=\mu^2=m_\tau^2$.}
\end{figure}

When considering the purely perturbative interpretation of Eqs.~(\ref{eq:BorelCIPT}) and (\ref{eq:BorelFOPT}) (i.e.\ the {\it truncated} Taylor series in u or $\bar u$, which is always an analytic function) and expanding either in $\alpha_s(s_0)$ or $\alpha_s(-s)$ {\it prior to the contour integration}, both expressions lead to both the FOPT and CIPT moment series and are equivalent. Furthermore, both Borel representations are formally related through the {\it complex-valued} change of variables $u=\bar u \,\alpha_s(-s)/\alpha_s(s_0)=\bar u\, a(-x)/a_0$. If the full Borel function $B[\hat D](u)$ were also an analytic function, this change of variables would be sufficient to prove that both Borel representations are equivalent. However,  Eqs.~(\ref{eq:BorelCIPT}) and (\ref{eq:BorelFOPT}) are \textbf{not equivalent and lead to different Borel sums due to the presence of non-analytic IR renormalons in the Euclidean Borel function $B[\hat D](u)$}. Consider a generic IR renormalon contribution of the form $\frac{1}{(p-u)^\gamma}$, which leads to a cut along the positive real $u$-axis starting at $u=p$, see the left panel in Fig.~\ref{fig:ucontour} showing the complex $u$-plane. This makes the FOPT Borel representation~(\ref{eq:BorelFOPT}) only well-defined with an additional prescription on the $u$-integration is imposed. The most common prescription used in the literature, the principal value (PV) prescription, is to take the average of the deformations above and below the cut (paths 1a and 1b). We have indicated it already in Eq.~(\ref{eq:BorelFOPT}). In contrast, the CIPT Borel representation does not require a prescription because
$\alpha_s(-s)/\alpha_s(s_0)$ is complex along the contour integration over $s$ as long as ${\rm Im}[s]\neq 0$. From the perspective of the $u$-integration in the FOPT Borel representation, the $\bar u$ integration in the CIPT Borel representation never touches the cut and proceeds either entirely above (shown as path 2 for ${\rm Im}[x]>0$) or below it. 
Figure~\ref{fig:ucontour} also illustrates that the difference in the FOPT and CIPT Borel sums, called the \textbf{asymptotic separation}, arises from closing path 2 with either paths 1a or paths 1b at positive real infinity. For ${\rm Im}[x]>0$, the situation displayed in the figure, it arises from closing path 2 with path 1a.

\section{Essential Comments and the Asymptotic Separation}
\label{sec:implications}

The form of the Borel representation of the CIPT spectral moments in Eq.~(\ref{eq:BorelCIPT}) is imperative when deriving the Borel function explicitly from the CIPT series terms. 
The analytic form of the CIPT Borel representation bears a number of novel and quite subtle properties which we briefly discuss in the following and which are important for the numerical computation of the asymptotic separation. In the following we consider a generic IR renormalon term in the Euclidan Adler functions Borel function of the form $B^{\rm IR}_{\hat D,p,\gamma}(u) = \frac{1}{(p-u)^\gamma}$.

\subsection{Form of the Contour Integration and CIPT OPE Corrections}

For the FOPT Borel representation the choice of the complex $x$-integration path ${{\cal C}_x}$ is arbitrary as long as it is ensured that the coupling $a(-x)$ stays in the perturbative region. For the CIPT Borel representation an additional restriction arises because the coupling affects the analytic properties of the CIPT Borel function~(\ref{eq:BorelCIPTTaylor}). Let us consider the contribution to the CIPT Borel representation due to the generic IR renormalon $B^{\rm IR}_{\hat D,p,\gamma}(u)$ and for $W(x)=(-x)^m$:
\begin{eqnarray}
\label{eq:BorelCIPT3}
\delta_{\{(-x)^m,p,\gamma\},{\rm Borel}}^{(0),{\rm CIPT}}(s_0) & = &
\textstyle
\int_0^\infty \!\! {\rm d} \bar u \, 
\frac{1}{2\pi i}\,\ointctrclockwise_{{\cal C}_x} \frac{{\rm d}x}{x} \,(-x)^m \,
\big({\textstyle \frac{a(-x)}{a_0}}\big)\,
\frac{e^{-\frac{\bar u}{a_0}}}{\big(p-\frac{a(-x)}{a_0}\bar u\big)^\gamma}\,.
\end{eqnarray}
Apart from the Landau pole and the cut along the positive real $x$ axis contained in the coupling $a(-x)$, there is an additional cut in the $x$-plane for real values of $x$ with
$\alpha_s(-x s_0) \ge p\alpha_s(s_0)/\bar u$. In the large-$\beta_0$ approximation this is equivalent to $x \ge \tilde x(\bar u)\equiv- e^{(\bar u-p)/p a_0} = -(\frac{\Lambda_{\rm QCD}^2}{s_0})^{(p-\bar u)/p}$. 
The value of $\bar x(u)$ is negative so that it affects the possible choices for the path ${{\cal C}_x}$. For $\bar u<p$ the cut is still within the circular path $|x|=1$. For $\bar u>p$ it is not, so that we have to deform the integration path ${\cal C}_x$ further into the negative real $x$-plane such that it crosses the real negative axis at a value below $\tilde x(\bar u)$. This property entails that, when $\bar u\to\infty$, the allowed region where the path can cross the negative real $x$-axis is shifted towards negative infinity. Furthermore, when the $\bar u$ integration is carried out first, this cut stretches to minus real infinity, so that the contour ${\cal C}_x$ must be deformed to minus negative real infinity as well. This additional cut is an essential issue when one attempts to apply the analytic structure of Eq.~(\ref{eq:BorelCIPT}) for a calculation of the Borel sum to the expansion of complex-valued (non-Euclidean) Adler function of Eq.~(\ref{eq:AdlerseriesCIPT}), i.e.\ when discussing its form without the contour integration (see Refs.~\cite{Hoang:2020mkw,Hoang:2021nlz} for such an analysis). 
The analytic form of the CIPT Borel representation implies that this Borel sum has a cut along the Euclidean axis that is  power-suppressed by a factor $e^{-\frac{p}{a(-x)}}\sim(\frac{\Lambda_{\rm QCD}^2}{-s})^p$. Since this cut is unphysical, one must conclude that the associated OPE corrections to the Adler function (which does not have such a cut at the hadron level) cannot have the standard form discussed in Sec.~\ref{sec:primer}.
This implies that the OPE corrections that need to be added to the CIPT spectral function moments differ from those of the FOPT moments and, furthermore, cannot be computed from the standard form of the Adler function's OPE corrections. We note that this conclusion is not imperative at this point since the contour integration is an integral part in the derivation of the form of Eq.~(\ref{eq:BorelCIPT}), 
but we believe that it is the correct one.\footnote{The unphysical cut may be taken as a formal reason to dismiss the form of Eq.~(\ref{eq:BorelCIPT}) and all its implications. However, because the cut is power suppressed, it can be compensated by OPE corrections that do not have standard form or maybe even have a connection to duality violating effect. So there is no contradiction. We believe that there is sufficient evidence that supports the view that the Borel representation of  Eq.~(\ref{eq:BorelCIPT}) should be taken seriously.} 

\subsection{FOPT and CIPT OPE Corrections are indeed different}
\label{sec:different}

The statement that the OPE corrections to the CIPT spectral function moments differ from those of the moments computed with FOPT and, furthermore, do not have standard form is is quite intriguing and not easy to accept. It implies that phenomenological analyses within the CIPT approach may be subject to a yet unquantified additional uncertainty concerning the treatment of the OPE corrections.\footnote{The size of this uncertainty is only sizeable if the normalization of the gluon condensate renormalon in full QCD is sizeable as well.} It is therefore worth to spend some time to discuss it further, having in mind the statement we made earlier on the suppression of the gluon condensate renormalon for spectral function moments $W(x)$ without a quadratic term $x^2$. 
In the large-$\beta_0$ approximation, where the gluon condensate renormalon structure in $B[\hat D](u)$ has the form $\frac{1}{2-u}$, it is straightforward to see this suppression when carrying out the $x$ contour integration for the FOPT Borel representation for this renormalon structure~\cite{Ball:1995ni}:
\begin{eqnarray}
\label{eq:BorelFOPT2} 
\delta_{\{(-x)^m,2,1\},{\rm Borel}}^{(0),{\rm FOPT}}(s_0) & = & 
\textstyle
{\rm PV}  \int_0^\infty \!\! {\rm d}u \, 
\frac{1}{2-u}\,
\frac{1}{2\pi i}\,\ointctrclockwise_{|x|=1} \frac{{\rm d}x}{x} \,(-x)^m \,
e^{-u\ln(-x)}\, e^{-\frac{u}{a_0}}\\ \nonumber
& = & 
\textstyle
{\rm PV}  \int_0^\infty \!\! {\rm d}u \, 
\,\frac{(-1)^m \sin(u\pi)}{\pi(u-m)}\,  \frac{1}{2-u}\,e^{-\frac{u}{a_0}}	\,.
\end{eqnarray} 
For $m\neq 2$ the renormalon pole at $u=2$ is completely eliminated, the Borel function for the FOPT moment becomes analytic in the entire $u$-plane and the PV prescription can be dropped. This cancellation is accompanied by two more facts, namely that (i) the associated FOPT series is convergent (see the red dots in the right panel of Fig.~\ref{fig:ucontour} for $W(x)=1$) and that (ii) the gluon condensate correction~(\ref{eq:gluoncondensate}) vanishes in the $x$-contour integration since the residue is zero. The associated CIPT series (blue dots), however, is not convergent.\footnote{It is intriguing that this fact has apparently never been noticed in the literature prior to our work.} The CIPT Borel representation for $W(x)=1$ that arises from carrying out the $x$ contour integration has the form
\begin{eqnarray}
\label{eq:BorelCIPT4}
\delta_{\{1,2,1\},{\rm Borel}}^{(0),{\rm CIPT}}(s_0) & = & 
\textstyle
\int_0^\infty \!\! {\rm d} \bar u\, \left(\frac{-1}{2a_0}\right) Q\left(1,0,\frac{2-\bar u}{2a_0}\right) \,  e^{-\frac{\bar u}{a_0}}\,,
\end{eqnarray}
with $Q(1,0,\rho) = \frac{i}{2\pi}\,[\ln(\rho+i\pi) -  \ln(\rho-i\pi) ]$. The $\bar u$-integral along the positive real axis does again not need a prescription, but the Borel function has cuts located parallel to the real $\bar u$ axis starting at distance $2|1+i a_0\pi|$ to the origin. These cuts signal that the underlying series is not convergent as can be clearly seen in the figure. For a renormalon cut $\frac{1}{(p-u)^\gamma}$ the distance is $p|1+i a_0\pi|$, see Refs.~\cite{Hoang:2020mkw,Hoang:2021nlz} for details and formulae for all cases. The non-convergence of the CIPT series and the uncancelled cuts, imply that (in the large-$\beta_0$ approximation) that the CIPT series requires a finite OPE correction. Since the standard gluon condensate OPE corrections vanishes, the required OPE corrections cannot have the standard form of Eq.~(\ref{eq:gluoncondensate}).

\subsection{Asymptotic Separation}

In the right panel of Fig.~\ref{fig:ucontour} also the FOPT and CIPT Borel sums for the series associated to the Borel function contribution $\frac{1}{2-u}$ for $W(x)=1$ are shown as the colored horizontal lines. They can be computed directly from Eqs.~(\ref{eq:BorelFOPT2}) and (\ref{eq:BorelCIPT4}).
The difference between the two is called \textbf{asymptotic separation} and clearly visible. The FOPT series clearly converges to its Borel sum, while the CIPT series approaches its Borel sum at intermediate order prior to divergence. In general, it is more convenient to calculate the asymptotic separation by doing first the Borel integration, applying the argumentation concerning closing the paths 1a and 1b with path 2 with respect to the IR renormalon cut we have mentioned at the end of Sec.~\ref{sec:Borelrepresentations}. This leads to 
\begin{eqnarray}
\label{eq:Sepa1}
\Delta(m,p,\gamma,s_0) \, \equiv \,
\delta_{\{(-x)^m,p,\gamma\},{\rm Borel}}^{(0),{\rm CIPT}}(s_0) \, - \,
\delta_{\{(-x)^m,p,\gamma\},{\rm Borel}}^{(0),{\rm FOPT}}(s_0) \nonumber \\[3mm]  \textstyle
=\,
\frac{1}{2 \Gamma(\gamma)} \,\ointctrclockwise_{{\cal C}_x} \frac{{\rm d}x}{x} \, (-x)^m \,
{\rm sig}[{\rm Im}[x]]\,(a(-x))^{1-\gamma}\,e^{-\frac{p}{a(-x)}}
\,. 
\end{eqnarray}
for the asymptotic separation for the generic renormalon structure $B^{\rm IR}_{\hat D,p,\gamma}(u)$. For $m<p$, which covers all linear weight functions,\footnote{We recall that $B[\hat D](u)$ only contain non-analytic renormalon terms of the form $\frac{1}{(p-u)^\gamma}$ for $p=2,3,4\ldots$.} 
the asymptotic separation $\Delta$ can be computed in the prescribed way, but for $m>p$, the exponentially suppressed term $e^{-\frac{p}{a(-x)}}$ is beaten by the divergent $(-x)^m$ term when $x$ approaches $-\infty$ in the remaining contour integral. For this case one needs to determine $\Delta$ through analytic continuation, which boils down to the analytic formula determined for $m<p$. Details of this analytic continuation are given in Ref.~\cite{Hoang:2020mkw}. For $m=p$, we \textbf{define by hand} $\Delta(p,p,\gamma,s_0)  =  0$, because in this case the renormalon behavior is not suppressed in the moment series and the FOPT and CIPT series both exhibit an unstable and divergent character~\cite{Beneke:2012vb,Hoang:2020mkw} such that the discussion of a discrepancy between them is irrelevant from the purely practical point of view. In the large-$\beta_0$ approximation the analytic formulae are quite simple and read ($e^{-\frac{p}{a_0}} = (\frac{\Lambda_{\rm QCD}^2}{s_0})^p$)
\begin{align}
\label{eq:Sepa4}
\textstyle
& \textstyle \Delta_{\beta_0}(m\neq p,p,1,s_0) =  \frac{ (-1)^{p-m}}{p-m}\,e^{-\frac{p}{a_0}}\,, \\
& \textstyle \Delta_{\beta_0}(m\neq p,p,2,s_0) = (-1)^{p-m}\,\bigg[ \frac{1}{(p-m)^2} +  \frac{1}{(p-m)a_0}\bigg]\,e^{-\frac{p}{a_0}}\,.
\end{align}
The analytic expressions in full QCD are more complicated and written down in  Ref.~\cite{Hoang:2020mkw}.

\begin{figure} 
	\centering
	\includegraphics[width= 0.49\textwidth]{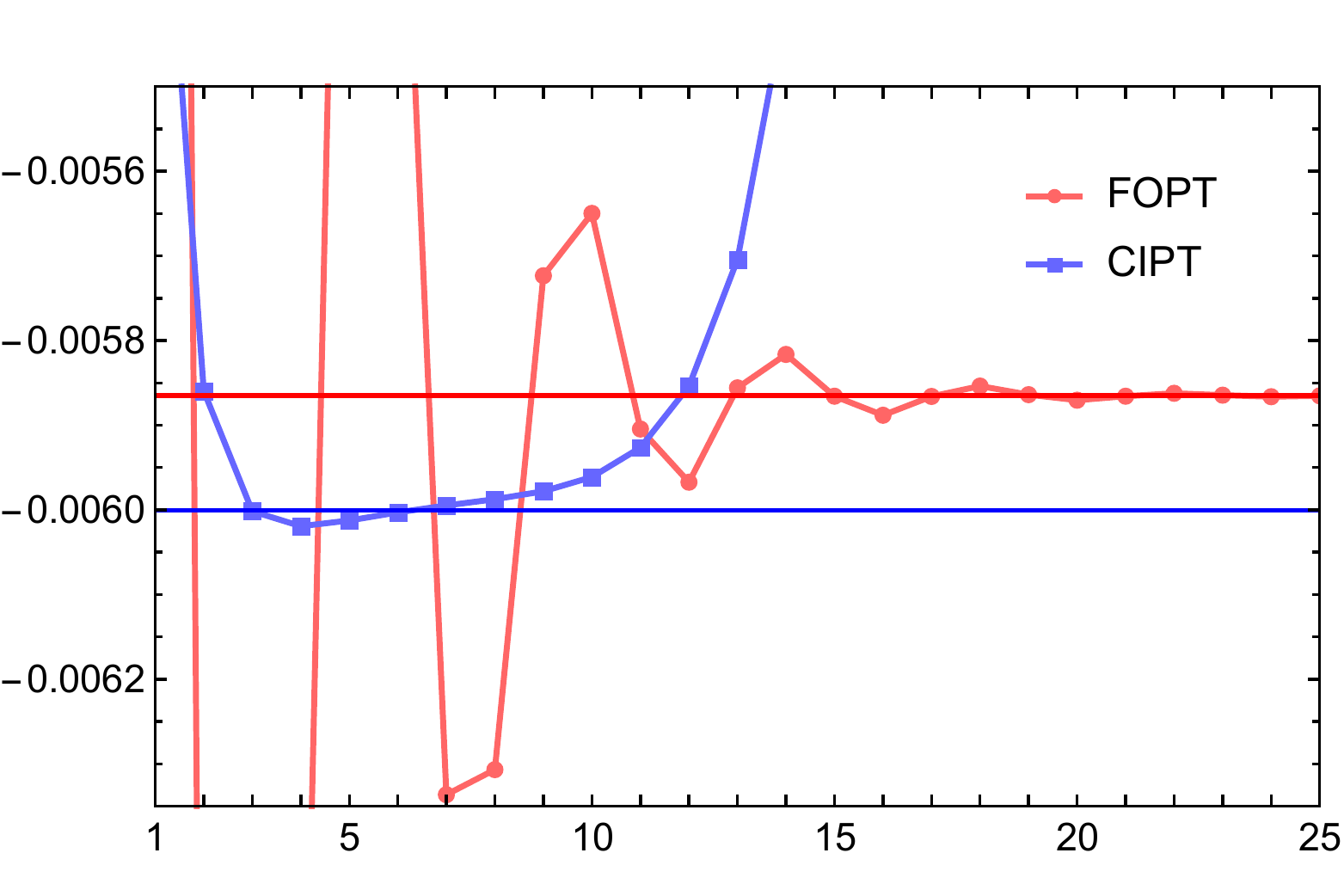}
	\includegraphics[width= 0.47\textwidth]{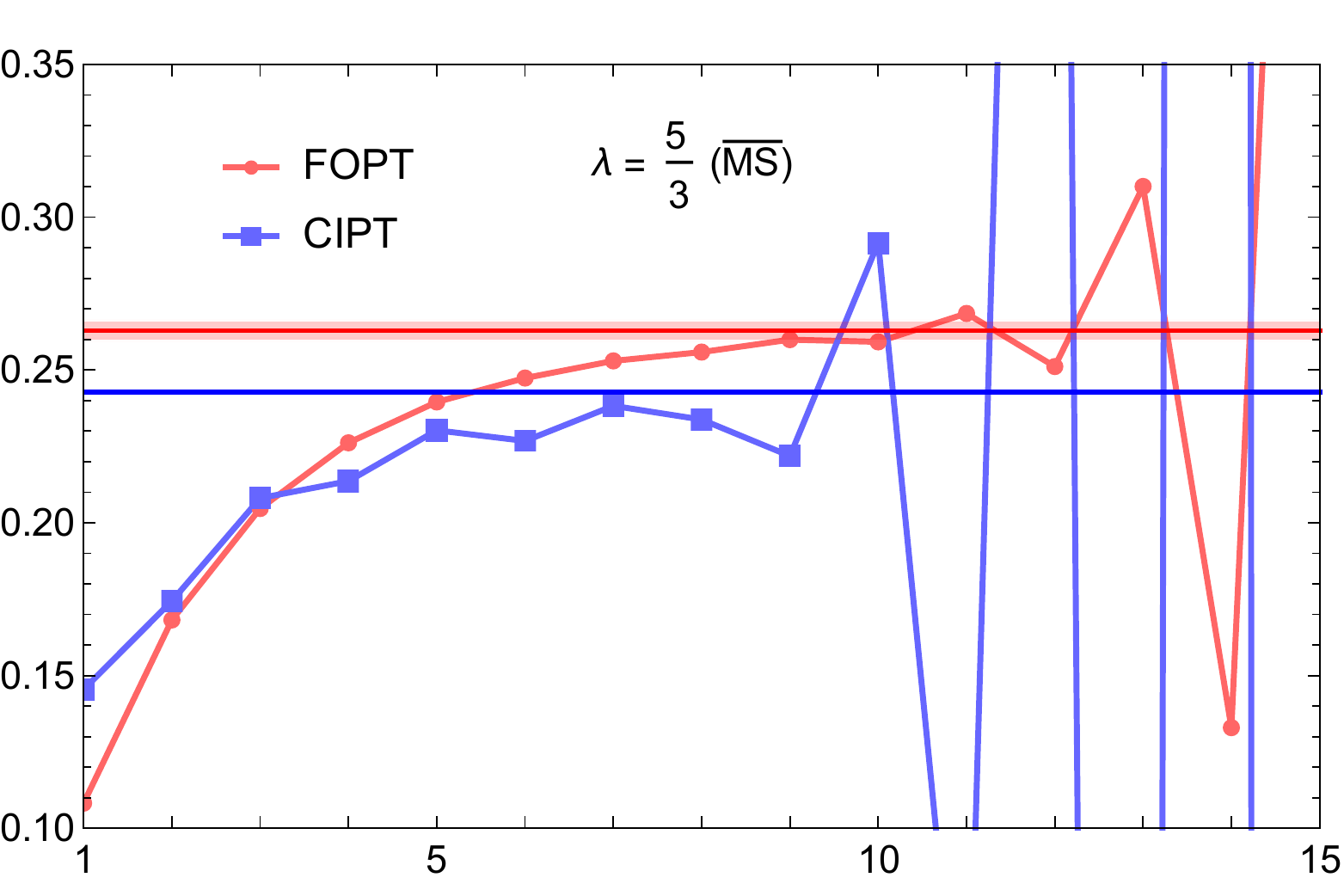}
	\caption{\label{fig:Rtau} \small
		Left: FOPT and CIPT spectral function moment series associated to the weight function $W(x)=(-x)^4$ and the Borel function $\frac{1}{2-u}$ in the large-$\beta_0$ approximation for $s_0=m_\tau^2$. Right: FOPT and CIPT Moment series $\delta^{(0),{\rm FOPT}}_{W_\tau}(m_\tau^2)$ for the total hadronic $\tau$ decay rate $R_\tau$ in the large-$\beta_0$ approximation. Horizontal lines represent the FOPT and CIPT Borel sums and the orange band in the right panel shows the FOPT Borel sum ambiguity. We used $\alpha_s(m_\tau^2)=0.34$ and $\mu^2=m_\tau^2$.}
\end{figure}

\subsection{Brief numerical Analysis}

We have already shown in the right panel of Fig.~\ref{fig:ucontour}  that the asymptotic separation describes the disparity in the behavior of FOPT and CIPT spectral function moments series very accurately for $W(x)=1$ and a $p=2$ simple pole IR renormalon in the large-$\beta_0$ approximation. The left panel of Fig.~\ref{fig:Rtau} shows the corresponding case for $W(x)=(-x)^4$, where the analytic continuation is mandatory to obtain the result for the asymptotic separation. Here the FOPT series is again convergent (in contrast to the CIPT series), and the description of the disparity in the behavior of both series is again very accurately described by the asymptotic separation. This excellent description can be easily checked for any IR renormalon and any monomial weight function $W(x)=(-x)^m$ also in full QCD and we refer to Ref.~\cite{Hoang:2020mkw} for details. 

We conclude with showing in the right panel of  Fig.~\ref{fig:Rtau} the FOPT and CIPT series for the normalized total hadronic $\tau$ decay rate $R_\tau$, where the weight function $W_\tau(x)$ is a linear combination of several monomials (see text below Eq.~(\ref{eq:deltadef})) using the full Borel function $B[\hat D](u)$ in the large-$\beta_0$ approximation~\cite{Broadhurst:1992si}, see e.g.\ Eq.~(12) in Ref.~\cite{Hoang:2021nlz} for the expression. Since  in the full Borel function IR and UV renormalon poles are located at all integer values along the real $u$-axis (except for $u=0,1$), the FOPT and the CIPT series are both asymptotic (and non-convergent). Both series are shown for the same $\alpha_s$ value. The oscillating structures visible in both series arise from the influence of UV renormalons which are associated to a sign-alternating increase of the series coefficients. The impact of these UV renormalons is, however, very small at intermediate orders below 9 so that we can observe the impact of the IR renormalons.\footnote{The leading UV renormalon pole is located at $p=-1$ and should in principle dominate over the impact of the gluon condensate renormalon pole at $p=2$ already at very low orders. However, it happens that in the $\overline{\rm MS}$ scheme the normalization of these UV renormalons is strongly suppressed compared to the IR renormalons.}  We can clearly see the disparity between the FOPT and CIPT series (around orders 5 to 8), which is the reason why $\alpha_s$ values based on CIPT analyses tend to be larger than for FOPT-based analyses (once the same OPE corrections are used in both approaches). 
The FOPT and CIPT Borel sums are again indicated by the colored horizontal lines and we have also displayed as the light orange band the standard estimate for the ambiguity of the FOPT Borel sum in the PV prescription, which is defined as the difference from using paths 1a and 1b in Fig.~\ref{fig:ucontour} (left panel) times a factor $\frac{1}{\pi}$.
We again see that the asymptotic separation describes the disparity between the FOPT and CIPT series very well, and we also observe that the asymptotic separation is substantially larger than the FOPT Borel sum ambiguity (even if we would not include the ad hoc suppression factor $\frac{1}{\pi}$). Interestingly, $99.8\%$ of the numerical value of the asymptotic separation comes from the gluon condensate renormalon. This happens because the Borel function $B[\hat D](u)$ of the Euclidean Adler function in the large-$\beta_0$ approximation contains a gluon condensate IR renormalon cut with a sizeable normalization and because the contribution of IR renormalons with  $p\ge 3$ is strongly power-suppressed by additional factors of $\Lambda_{\rm QCD}^2/s_0$, see Eq.~(\ref{eq:Sepa4}). From a practical point of view, only the $p=2$ gluon condensate renormalon is relevant when considering the implications of the asymptotic separation. 

\section{Conclusions}

In this talk we have shown that the Borel representations of the FOPT and CIPT $\tau$ hadronic spectral function moments have different Borel representations. The CIPT Borel representation is new and has novel and subtle features.
In the presence of IR renormalons the different analytic properties lead to a difference in their Borel sums, called the \textbf{asymptotic separation}. 
While the FOPT Borel representation has been known before, the CIPT Borel representation is new and its structure provides the implication that the OPE corrections that need to be added to the CIPT moments differ from those of the FOPT approach and do furthermore not have standard form. From a numerical point of view, the asymptotic separation and its implications are practially relevant only if the gluon condensate IR renormalon has a substantial normalization. This is so in the large-$\beta_0$ approximation, where the asymptotic separation nicely describes the disparity between the FOPT and CIPT spectral function moment series.

\section*{Acknowledgements}
We thank 
Matthias Jamin and Diogo Boito for discussions. This work was supported in part by the FWF Austrian Science Fund under the Project No. P32383-N27.
We also acknowledge partial support by the FWF Austrian Science Fund under the Doctoral Program ``Particles and Interactions'' No.\ W1252-N27.

\bibliographystyle{SciPost_bibstyle}
\bibliography{sources.bib}

\nolinenumbers

\end{document}